\documentclass[epj]{svjour}
\usepackage{graphics}
\begin{document}
\title{NICA fixed target mode: soft jet studies in the relative 4-velocity space}

\author{V.A. Okorokov\inst{}
\thanks{e-mail: VAOkorokov@mephi.ru;~Okorokov@bnl.gov}%
}                     
\institute{National Research Nuclear University MEPhI (Moscow
Engineering Physics Institute), \\Kashirskoe Shosse 31, 115409
Moscow, Russian Federation}
\date{Received: 27 September 2015 / Revised version: 7 February 2016}
%
\abstract{Experimental results obtained by studying the properties
of soft jets in the 4-velocity space at $\sqrt{s} \sim 2-20$ GeV
are presented. The changes in the mean distance from the jet axis
to the jet particles, the mean kinetic energy of these particles,
and the cluster dimension in response to the growth of the
collision energy are consistent with the assumption that quark
degrees of freedom manifest themselves in processes of pion-jet
production at intermediate energies. The energy at which quark
degrees of freedom begin to manifest themselves experimentally in
the production of soft pion jets is estimated for the first time.
The estimated value of this energy is $2.8 \pm 0.6$ GeV. The
suggestions are made for future investigations on NICA.
} 
\maketitle
\section{Introduction} \label{sec:1}

At the present time, the decision of the problem of confinement
and the study of transition from meson-baryon degrees of freedom
to quark-gluon ones is one of the most important (and, at the same
time, most difficult) tasks of the world research programm in the
field of strong interactions. The nature of confinement of color
degrees of freedom (quark and gluons) and, correspondingly, the
possible phase transitions in strongly interacting matter is not completely
clear so far. Observation of hadron jets at high energies
is one of the most important and evident experimental
manifestation of quark-gluon degrees of freedom. At present, an
open question is where is the low boundary on energy starting with
which the color degrees of freedom should be taken into account
for the description of processes of multiparticle production.
Obviously, the jet structure of events displays itself more
clearly at high initial energies ($\sqrt{s}$) than at intermediate
ones. But in spite of this feature it seems that the application
of collective characteristics of multiparticle final state can be
useful in the collision energy domain $\sqrt{s} \simeq 2-20$ GeV both
for the study of transition from the predominance of meson-baryon
degrees of freedom to the quark-gluon ones and for qualitative
estimation of low boundary (for initial energy) experimental
manifestation of quark degrees of freedom in soft jet production.
In various fields of physics the onset of manifestation of new
degrees of freedom and transition processes is accompanied by the
presence of self-affine and fractal properties in collective
effects. Therefore, the precise measurements of collective and
geometric (fractal-like) properties of soft pion jets in the
NICA energy domain can give a new important information
about hadronization mechanisms, behavior of quantum systems in the
nonperturbative region and transition to manifestation of quark
degrees of freedom in collective phenomena.

\section{Method and variables} \label{sec:2}

Traditional collective characteristics used for the study of the event
shape \cite{Okorokov-IJMPA-27-1250037-2012} are not
relativistically invariant. This introduces some additional
kinematic uncertainties, for example, in the choice of the center-of-mass
system for reactions with atomic nuclei. A relativistic-invariant
method was proposed in \cite{Baldin-DoklANUSSR-222-1064-1975} for
studying collective effects in the case where particle-beam
interaction with a target leads to the formation of a
multiparticle final state in the reaction $\mbox{b}+\mbox{t} \to
1+2+\dots$ Special features of this method in the case of the
production of two jets were considered in detail in
\cite{Okorokov-YaF-57-2225-1994,Okorokov-YaF-62-1787-1999,Okorokov-YaF-73-2016-2010,Okorokov-IJMPA-28-1350150-2013,Okorokov-YaF-78-445-2015}.
In that case, secondary particles refer to the region of target
(beam) fragmentation if $X^{k}_{\mbox{\footnotesize{t}}} \geq \!
(\leq) \tilde{X} \bigcap X^{k}_{\mbox{\footnotesize{b}}} \leq \!
(\geq) \tilde{X}$, where
$X^{k}_{\mbox{\footnotesize{p}}}=[m_{k}(U_{k}U_{\mbox{\footnotesize{r}}}
)][m_{\mbox{\footnotesize{p}}}(U_{\mbox{\footnotesize{p}}}U_{\mbox{\footnotesize{r}}})]^{-1}$,
$\mbox{p}, \mbox{r}=\mbox{t}, \mbox{b}$; and $\mbox{p} \ne
\mbox{r}$; $m_{k}$ is the mass of the
$k^{\,\mbox{\footnotesize{th}}}$ secondary particle;
$m_{\mbox{\footnotesize{t\,/\,b}}}$ is the mass of the target/beam
particles; $U_{k}=P_{k}/m_{k}$ is the 4-velocity; $k=\mbox{t},
\mbox {b}, 1, 2, \dots$; and $\tilde{X}= 0.1-0.2$ is some boundary
value which is determined empirically. The basic quantities which
the probability distributions (cross sections) depend upon are
non-dimensional positive relativistic invariant quantities
$b_{ik}=-(U_{i}-U_{k})^{2}$, where $i, k=\mbox{t},
\mbox{b},1,2,...$ \cite{Baldin-DoklANUSSR-222-1064-1975}. As seen
the observables $b_{ik}$ mean the squares of relative distances in
the four-velocities space. The comparison of this method for
distinguishing of some particle groups in the space of
four-dimensional velocities with other present non-invariant
(traditional) methods allows to name these separate groups as jets
\cite{Baldin-YaF-44-1209-1986}. One of the most important
observables of this approach is defined as
$b_{k}=-\left(V-U_{k}\right)^{2},~k=\mbox{t}, \mbox{b}, 1, 2,
\dots$, where $V=U_{J}/|U_{J}|$, $U_{J}=\sum^{N}_{i=1} U_{i}$ and
$N$ is the number of particles in the considered fragmentation region
which satisfies all cuts and is involved in the analysis
\cite{Baldin-DoklANUSSR-222-1064-1975,Baldin-Lektcii-43-P1-87-912-1987}.
The quantity $b_{k}$ is the square of the distance of the
$k^{\,\mbox{\footnotesize{th}}}$ particle from the jet axis $V$ in
the space of $U_{k}$. The ``temperature" defined as the mean
kinetic energy of particles in the jet rest frame, $\langle
T_{k}\rangle$, is estimated on the basis of the invariant
functions $F(b_{k})$ that, for pion jets, have the form
\cite{Okorokov-YaF-62-1787-1999} $F(b_{k}) = (\varepsilon / N)\,dN
/ db_{k}$, where $\varepsilon \equiv 4 /
[m_{\pi}^{2}b_{k}\sqrt{1+4/b_{k}}]$, and which characterize the
invariant cross section \cite{Baldin-Lektcii-43-P1-87-912-1987}.
In \cite{Okorokov-NSMEPhI-218-2000}, it was proposed to study the
geometric properties of jets in the $U_{k}$ space with the aid of
the cluster dimension $D$ defined on the basis of the relation
between the number of particles in the jet being considered,
$N(b_{k})$, and its radius: $N\left(b_{k}\right) \propto
b_{k}^{D/2}$. Non-integer value of the cluster dimension can be
considered as characteristic signature of manifestation of
fractal-like properties \cite{Fomenko-book-1998}. For most complex
distributions the multi-fractal structure can appear and the
cluster dimension be a function of jet radius for such case:
$D=D(b_{k})$. Thus $D$ is the qualitative parameter reflecting
the features of particle distribution in phase space. The set of
observables $\mathcal{G} \equiv
\{\mathcal{G}_{i}\}_{i=1}^{3}=\{\langle b_{k}\rangle, \langle
T_{k}\rangle, D\}$ characterizing the geometry and dynamics of the
final-state production in the 4-velocity space is under consideration and is
proposed for future investigation on NICA.

\section{Recent results} \label{sec:3}

\begin{figure*}
\vspace*{1cm}
\begin{center}
\resizebox{0.81\textwidth}{!}{%
\includegraphics{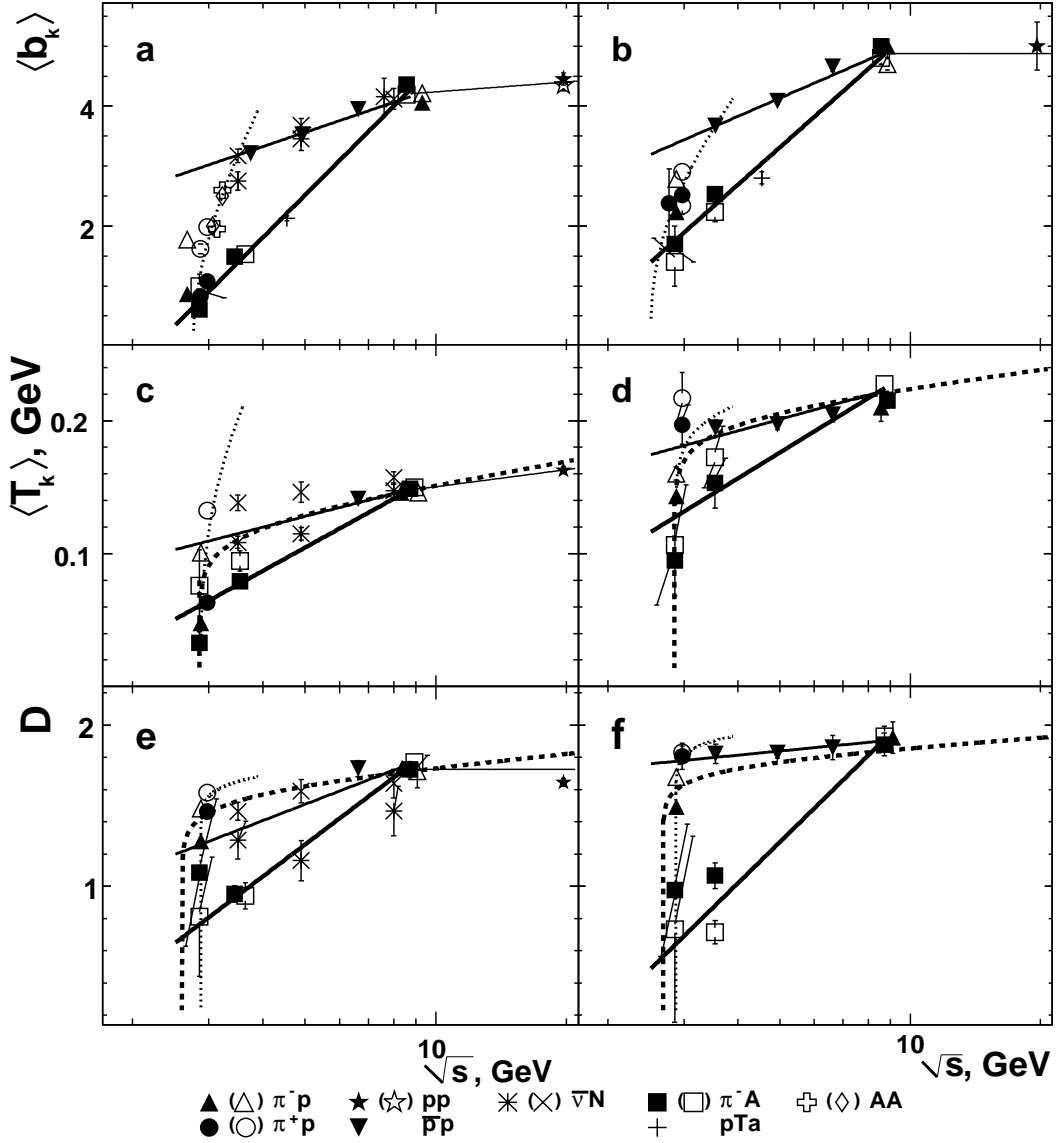}}
\end{center}
\caption{Dependence of parameters from set $\mathcal{G}$ on
$\sqrt{s}$ at $\tilde{X}=0.1$ (a, c, e) and 0.2 (b, d, f) in the
region of target (beam) fragmentation
\cite{Okorokov-YaF-78-445-2015}. The dotted line corresponds to
the approximation of data in the region of $\sqrt{s} < 4$ GeV by
the function (\ref{eq:1}), while the dashed lines in Figs. c -- f
represent the approximation of all available experimental data by
the function (\ref{eq:1}). The solid lines stand for the results
obtained by fitting the logarithmic function (\ref{eq:2}) to a
global sample of $hh$ and $\bar{\nu}N$ reactions for $4 \leq
\sqrt{s} < 9$ GeV (line with moderate thickness), data on $hA$
reactions (thick line), and data on $hh$ and $hA$ interactions for
$\sqrt{s} > 8$ GeV (thin line).} \label{fig:1}
\end{figure*}

Figure \ref{fig:1} gives the parameters from the set $\mathcal{G}$
versus $\sqrt{s}$ for various interactions at $\tilde{X}=0.1$ (a,
c, e) and $\tilde{X}=0.2$ (b, d, f). The experimental-data array
from
\cite{Okorokov-IJMPA-28-1350150-2013,Okorokov-YaF-78-445-2015} is
under study and various samples of these data are approximated
by the functions
\begin{equation}
\mathcal{G}_{i}=a_{1}\bigl(\sqrt{s/s_{0}}-a_{2}\bigr)^{a_{3}},~~\sqrt{s/s_{0}}
\geq a_{2};\label{eq:1}
\end{equation}
\begin{equation}
\hspace{-2.55cm}\mathcal{G}_{i}=a_{1}+a_{2}\ln(s/s_{0}).
\label{eq:2}
\end{equation}
where $i=1-3$ and $s_{0}=1$ GeV$^{2}$. One can see a qualitatively
similar character of the energy dependence for all parameters from
the set $\mathcal{G}$ at the respective values of $\tilde{X}$.
Figure \ref{fig:1} shows that, for any value of $i$, the
dependencies $\mathcal{G}_{i}(\sqrt{s})$ exhibit a change in
behavior in the region around $\sqrt{s} \simeq 3$ GeV for all
interaction types, with the exception of $hA$, at any value of
$\tilde{X}$, and this confirms the hypothesis put forth in
\cite{Okorokov-YaF-62-1787-1999} that dynamical interaction
regimes undergo a change for $\sqrt{s} < 3-4$ GeV.

In the region of $\sqrt{s} < 4$ GeV, experimental dependencies
$\mathcal{G}_{i}(\sqrt{s})$, $i=1-3$, for all interaction types,
with the exception of $hA$, are approximated by the function in
(\ref{eq:1}). The results of samples that combine the
fragmentation regions are shown by dotted lines. One can see that
the function in (\ref{eq:1}) agrees qualitatively with
experimental data at all of the $\tilde{X}$ values considered
here. The parameter $a_{2}$ in (\ref{eq:1}) can be put in
correspondence with the energy $\sqrt{s_{c}}$ at which
quark--gluon degrees of freedom begin to manifest themselves in
the production of soft pion jets. The values of $\sqrt{s_{c}}$ for
the members of $\mathcal{G}$ are given in the Table \ref{tab:1}.
The estimates of $\sqrt{s_{c}}$ for $\langle T_{k}\rangle$ and $D$
agree well with each other but exceed the values of $\sqrt{s_{c}}$
obtained earlier in \cite{Okorokov-YaF-73-2016-2010} on the basis
of the $\langle b_{k}\rangle (\sqrt{s})$ dependence. This
difference may be due both to physical reasons that lead to a
sharper growth of $\langle T_{k}\rangle$ and $D$ in relation to
$\langle b_{k}\rangle$ and to a smaller size of the
experimental-data samples in the case of the first two parameters.
In the energy region being considered, an approximation by the
function in (\ref{eq:1}) is also constructed for individual
fragmentation regions. There is a substantial improvement of the
quality of fits for any fragmentation region and any value of
$\tilde{X}$ ($\chi^{2}/\mbox{ndf} \sim 3-5$). Statistically
acceptable values of $\chi^{2}/\mbox{ndf}$ could be obtained for
45\% of the samples
\cite{Okorokov-IJMPA-28-1350150-2013,Okorokov-YaF-78-445-2015}.
Numerical values found for $\sqrt{s_{c}}$ from a fit to
$\mathcal{G}_{i}(\sqrt{s})$ for $\sqrt{s} < 4$ GeV are summarized
in the Table \ref{tab:1}, the values corresponded to the fits of
$\langle b_{k}\rangle$ energy dependence being taken from
\cite{Okorokov-YaF-73-2016-2010}. The values of $\sqrt{s_{c}}$ for
the remaining two parameters from $\mathcal{G}$ are presented in
the first lines for $\langle T_{k}\rangle$ and $D$. Taking into
account the sizes of samples of the available experimental data and
the behavior of $\langle T_{k}\rangle(\sqrt{s})$ and
$D(\sqrt{s})$, we can approximate the dependencies in question by
the function in (\ref{eq:1}) for $\sqrt{s} > 2$ GeV. A fit was
constructed both for global samples and for the regions of target
and beam fragmentation. For the first case, the fitted results are
shown by the dashed curves in Figs. \ref{fig:1}c -- f. For
$\langle T_{k}\rangle$, agreement between the approximation by the
function in (\ref{eq:1}) and experimental data is only qualitative
in the case of $\tilde{X}=0.1$ (see Fig. \ref{fig:1}c), since the
fit quality is substantially poorer in that case than in the case
of $\tilde{X}=0.2$ (see Fig. \ref{fig:1}d). For $D$, the fit
quality is indicative of only qualitative agreement between the
function in (\ref{eq:1}) and experimental data at either value of
$\tilde{X}$ (see Figs. \ref{fig:1}e, f)
\cite{Okorokov-IJMPA-28-1350150-2013,Okorokov-YaF-78-445-2015}. In
the Table \ref{tab:1}, the values obtained for $\sqrt{s_{c}}$ from
a fit in the region of $\sqrt{s} > 2$ GeV for global samples and
for individual samples in the target- and beam fragmentation
regions are given in the second rows for $\langle T_{k}\rangle$
and $D$. The experimental results for the traditional definition
of the thermal freeze-out temperature ($T$) obtained for $pp$
collisions at $\sqrt{s} \approx 6-18$ GeV
\cite{Abgrall-EPJC-74-2794-2014} agrees quite reasonably with
$\langle T_{k}\rangle (\sqrt{s})$ at soft $\tilde{X}=0.1$ (see
Fig. \ref{fig:1}c) in both the functional behavior and the
magnitude. As consequence, the thermal freeze-out temperatures
\cite{Abgrall-EPJC-74-2794-2014} are some smaller than the mean
``temperatures" of pions in jets at hard cut $\tilde{X}=0.2$ and
close energies. It should be noted $T(\sqrt{s})$ calculated
within the framework of the self-similarity approach assumes the sharp
decrease at $\sqrt{s} < 5$ GeV and is almost flat at larger energies
\cite{Artemenkov-IJMPA-30-1550127-2015}. Such behavior of
$T(\sqrt{s})$ agrees well with the features of $\langle
T_{k}\rangle (\sqrt{s})$ (Fig. \ref{fig:1}c, d) and confirms the
hypothesis with regard to the changing of the dynamical regimes of
multiparticle production at $\sqrt{s} \sim 5$ GeV
\cite{Okorokov-YaF-62-1787-1999}.
\begin{table*}
\caption{Values of $\sqrt{s_{c}}$ (in GeV units) at the boundary
values of $\tilde{X}=0.1$ and 0.2 \cite{Okorokov-YaF-78-445-2015}}
\label{tab:1}
\begin{center}
\begin{tabular}{ccccccc} \hline\noalign{\smallskip}
\multicolumn{1}{c}{Parameter} & \multicolumn{2}{c}{Global sample}
& \multicolumn{2}{c}{Target fragmentation}
& \multicolumn{2}{c}{Beam fragmentation} \\
\cline{2-7}
from set $\mathcal{G}$& 0.1 & 0.2 & 0.1 & 0.2 & 0.1 & 0.2 \rule{0pt}{8pt}\\
\hline
$\langle b_{k}\rangle$ & $2.76 \pm 0.01$  & $2.51 \pm 0.03$  &$2.82 \pm 0.02$   & $2.46 \pm 0.04$   & $2.43 \pm 0.04$   & $2.5 \pm 0.3$ \rule{0pt}{10pt}\\
$\langle T_{k}\rangle$ & $2.877 \pm 0.001$& $2.865 \pm 0.007$&$2.75 \pm 0.04$   & $2.865 \pm 0.007$ & $2.877 \pm 0.001$ &  -- \\
                       & $2.854 \pm 0.001$& $2.854 \pm 0.001$&$2.853 \pm 0.001$ & $2.854 \pm 0.001$ & $2.877 \pm 0.001$ & $2.854 \pm 0.001$\\
$D$                    & $2.875 \pm 0.001$& $2.875 \pm 0.001$&$2.875 \pm 0.004$ & $2.877 \pm 0.001$ & $2.78 \pm 0.04$   &  -- \\
                       & $2.60 \pm 0.03$  & $2.68 \pm 0.03$  &$2.854 \pm 0.002$ & $2.858 \pm 0.001$ & $2.9 \pm 0.4$     & $2.874 \pm 0.001$\\
\noalign{\smallskip}\hline
\end{tabular}
\end{center}
\end{table*}

One can see that the values of $\sqrt{s_{c}}$ are reasonably
consistent for the different $\tilde{X}$ values and fragmentation
regions. Thus, this investigation extended for the whole set
$\mathcal{G}$ renders the results obtained earlier in
\cite{Okorokov-YaF-62-1787-1999,Okorokov-YaF-73-2016-2010} more
reliable and furnishes an additional argument in support of the
hypothesis that a changeover of dynamical interaction regimes
occurs because of the onset of the experimental manifestation of
quark degrees of freedom in the production of soft pion jets at
$\sqrt{s} \sim 3$ GeV. This entails the respective transition from
the description of the processes in question in terms of
baryon--meson degrees of freedom to the use of quark--gluon
degrees of freedom. The lower boundary of the energy corresponding
to the onset of the experimental manifestation of quark degrees of
freedom in the production of soft pion jets was estimated
quantitatively for the first time. The interval in which the
estimates of this parameter are contained and which is matched
with the results for all of the collective features under
consideration from $\mathcal{G}$ (see Table \ref{tab:1}) is
$\sqrt{s_{c}}=2.43-2.90$ GeV. Treating the set of estimates of
$\sqrt{s_{c}}$ from the Table \ref{tab:1} as a sample of
independent measurements, one can obtain $\langle
\sqrt{s_{c}}\,\rangle = 2.78 \pm 0.14$ GeV. On the other hand,
allowance for the amplitude of changes in $\sqrt{s_{c}}$ within
one standard deviation (see Table \ref{tab:1}) yields the interval
$[(\sqrt{s_{c}})_{\scriptsize{\mbox{min}}};
(\sqrt{s_{c}})_{\scriptsize{\mbox{max}}}]$ for which the choice of
midpoint and mean deviation makes it possible to obtain an
estimate of $\langle \sqrt{s_{c}}\,\rangle=(2.8 \pm 0.6)$ GeV
\cite{Okorokov-IJMPA-28-1350150-2013,Okorokov-YaF-78-445-2015}. A
universal lower boundary for the manifestation of jet geometry for
final states was qualitatively estimated in
\cite{Okorokov-YaF-76-2013} at $\sqrt{s_{l}} \sim 3$ GeV for
multiparticle-production processes. This value agrees with
$\sqrt{s_{c}}$ with allowance for the errors. Possibly, the
parameters $\sqrt{s_{l}}$ and $\sqrt{s_{c}}$ characterize the same
physical effect -- the onset of experimental manifestations of
quark degrees of freedom in soft processes of the multiparticle
production of secondaries and, hence, a manifestation of the jet
structure of the event being considered. Thus, the results
obtained by using traditional and four-dimensional collective
variables agree reasonably and supplement each other.

Taking into account the results for $\langle
b_{k}\rangle(\sqrt{s})$ in the region of $\sqrt{s} > 3.5$ GeV
\cite{Okorokov-YaF-62-1787-1999,Okorokov-YaF-73-2016-2010} and the
behavior of $\langle T_{k}\rangle(\sqrt{s})$ and $D(\sqrt{s})$ in
the above range of $\sqrt{s}$, we approximated by the function
(\ref{eq:2}) the samples taken for $\langle T_{k}\rangle$  and $D$
and summed over the fragmentation regions. The numerical values of
the fit parameters and the respective detailed discussion are
presented in \cite{Okorokov-IJMPA-28-1350150-2013}. From Fig.
\ref{fig:1}, one can see that, for $3.5 < \sqrt{s} < 9$ GeV, the
values of $\mathcal{G}_{i}$, $i=1-3$, grow faster for $hA$
reactions than for a nucleon target. Analyzing the results
obtained for $\langle T_{k}\rangle$ and $D$ in the range of
$\sqrt{s} \sim 3-5$ GeV and taking into account large statistical
errors in $hA$ data, one can extend the conclusion concerning the
effect of nuclear matter on the properties of soft pion jets
\cite{Okorokov-YaF-57-2225-1994} to the whole set $\mathcal{G}$.
Approximations for $\langle T_{k}\rangle (\sqrt{s})$ and
$D(\sqrt{s})$ in the region of $\sqrt{s} > 8$ GeV are possible
only in the case of a soft limit on $\tilde{X}$ (see Figs.
\ref{fig:1}c, e). In view of the scantiness of the available data, we
cannot rule out definitively a weak logarithmic growth for $D$ in
accordance with Eq. (\ref{eq:2}). Thus, we see that, for all $i$,
the dependencies $\mathcal{G}_{i}(\sqrt{s})$ admit a universal
approximation in the form (\ref{eq:2}) for a broad class of
interactions at c.m. energies in the region of $\sqrt{s}
> 3.5$ GeV and for both $\tilde{X}$ values considered in the present
study.

\section{Further advancement} \label{sec:4}

It is noteworthy, that agreement of the results of investigation
of all parameters from the set $\mathcal{G}$ with one another and
the self-consistency of a global analysis of the properties of
soft pion jets in $U_{k}$ space renders the hypothesis of the
onset of experimental manifestation of quark degrees of freedom in
processes of the production of soft pion jets at $\sqrt{s} \sim 3$
GeV more plausible. However in order to test the above hypothesis
it would be reasonable to perform additional investigations in the
range of $\sqrt{s} \sim 2-20$ GeV. Furthermore the limited samples
of experimental results for $\langle T_{k}\rangle$ and $D$ and
significant errors especially for $\sqrt{s} \sim 3$ GeV allow a
semi-qualitative analysis only for these observables. Therefore
additional high-statistic investigations on the NICA with
various beams would be important for better understanding of the
dynamic and geometric features of the production of soft pion
jets.

The separation of various dynamics of jet production is a
difficult task especially in the non-perturbative region of
$\sqrt{s}$. The study of traditional collective observables at
such $\sqrt{s}$ allows the conclusion for event shape mostly
\cite{Okorokov-IJMPA-27-1250037-2012}. Within the approach in
question, the range $10^{-2} \leq b_{ik} \sim 1$ corresponds to
the transition from the domain of dominance of meson-baryon
degrees of freedom to the region where the internal structure of
colliding particles and, as a consequence, quark-gluon degrees of
freedom become essential in the processes of secondary particle
jet production\footnote{One needs to note that the boundary values
for $b_{ik}$ indicated above are qualitative phenomenological
estimations.}. The estimations for boundaries of various dynamic
domains in terms of $b_{ik}$ seem valid for $b_{k}$, i.e. when the
jet axis corresponds to the ``reference"
$i^{\,\mbox{\footnotesize{th}}}$ particle. That is why the
distributions on $b_{k}$ are usually studied. The approach under
discussion allows the separation, at least, at qualitative level
different dynamics of soft jet production. This advantage of the
relativistically invariant method is important for intermediate
$\sqrt{s}$ namely. Therefore approaches for the analysis of jet
production based on traditional collective characteristics and on
the set $\mathcal{G}$ are complementary to each other at
intermediate $\sqrt{s}$. Complete investigation of jet production
seems important especially at NICA for the energy region which
corresponds to the transition from the dominance of meson-baryon
degrees of freedom to quark-gluon ones.

Another suggestion is the study of the behavior of $\alpha_{S}$ --
the effective strong coupling constant -- in the deeply
non-perturbative region of $\sqrt{s}$. In the lowest order of the
renormalization group equation (RGE) the following relation was
derived $\alpha_{S} \propto 1/\ln b_{ik}$, where
$i=\mbox{t\,/\,b}$
\cite{Baldin-Lektcii-43-P1-87-912-1987,Baldin-NPA-434-695c-1985}.
The possible exact relation between $b_{k}$ and $\alpha_{S}$
requires an additional rigorous substantiation and careful
derivation. Here one can note only that jets consist of particles
with close masses (pions) in the present study and there is the
relation $\langle b_{k}\rangle = 2(\langle M_{J}\rangle / \langle
n_{J}\rangle m_{h}-1)$ between the geometry quantity of jet in $U_{k}$
space and mean invariant mass ($M_{J}$) of jet of hadrons with
equal masses $m_{h}$
\cite{Grishin-Book-1988,Okorokov-ISHEPP-154-2006}. Here $\langle
n_{J}\rangle$ is the mean multiplicity of particles inside the jet,
the averaging is taken over particles in the event and over events in the
sample in the l.h.s.; and over event ensemble in the r.h.s. The
appropriate choice of the energy scale $Q$ in the RGE is
non-trivial for hadronic processes especially in the deeply
non-perturbative region. By analogy with collider experiments
\cite{D0-PLB-718-56-2012,CMS-arXiv-1304.7498-2013} $\langle
M_{J}\rangle$ is chosen as $Q$ for some reaction at fixed
$\sqrt{s}$ within the framework of the method under study. Then
based on the \cite{PDG-2014} one can derive the relation
\begin{equation}
\alpha_{S} = (b_{0}\zeta)^{-1},~~\zeta \equiv
2\ln\bigl[(0.5\langle b_{k}\rangle+1)\langle n_{J}\rangle m_{h}
\Lambda^{-1}\bigr], \label{eq:My-new}
\end{equation}
where $b_{0}=(33-2n_{f})/(12\pi)$ is the one-loop $\beta$-function
coefficient, $n_{f}$ is the number of quark flavors considered as
light, $\Lambda$ is the QCD parameter. The (\ref{eq:My-new})
demonstrates possible relation between soft jet characteristic
$\langle b_{k}\rangle$ and strong coupling and gives indications on the
additional advantage for study in non-perturbative regime at
NICA. The estimations $\langle M_{J}\rangle \sim 1-2$ GeV obtained
by the method under consideration for $\pi p$ reactions at $\sqrt{s}
\sim 3$ GeV \cite{Okorokov-ISHEPP-154-2006} agree with the results
at $\sqrt{s} \sim 9$ GeV \cite{Badalyan-JINR-RC-1-27-1996}.
Therefore the suggestion seems reasonable with regard to the validity
of (\ref{eq:My-new}) in the NICA energy domain. It should be noted
that there is one measurement of $\alpha_{S}$ in the deeply
non-perturbative domain at $Q \sim 2$ GeV, namely from $\tau$
decay \cite{PDG-2014}. Thus the suggested equation
(\ref{eq:My-new}) can provide the important estimation of
$\alpha_{S}$ at $Q \sim \langle M_{J}\rangle$ on the order of few
GeV and verifies the validity the theoretical curve
$\alpha_{S}(Q)$ at low energy scales, i.e. it allows the
investigation of one of the fundamental properties of QCD in a
most difficult domain for theoretical description.

\section{Summary} \label{sec:5}

Summarizing the foregoing, we can draw the following conclusions.

The dependencies $\mathcal{G}_{i}(\sqrt{s})$, $i=1-3$, exhibit
qualitatively similar types of behavior, which admits a
description in terms of power-law function for $\sqrt{s} < 4$ GeV
and in terms of a logarithmic function for $\sqrt{s} > 3.5$ GeV.
The behavior of $\mathcal{G}_{i}(\sqrt{s})$ for all $i$ at
$\sqrt{s} \sim 3$ GeV is likely to be due to the onset of an
experimental manifestation of quark degrees of freedom in the
production of soft pion jets and the respective transition from
the description of the processes in question in terms of
baryon--meson degrees of freedom to the use of quark--gluon
degrees of freedom. A lower limit on the energy at which quark
degrees of freedom begin to manifest themselves in the production
of soft pion jets was estimated quantitatively for the first time.
The result is $\langle \sqrt{s_{c}}\,\rangle = (2.8 \pm 0.6)$ GeV.
The effect of nuclear matter on the dynamical and geometric
properties of soft pion jets in the 4-velocity space was found in
the range of $\sqrt{s} \sim 3-5$ GeV.

The investigations of collective and geometric (fractal)
properties in soft hadron and nuclear reactions at intermediate
energies with high statistics at NICA can provide important
progress for better a understanding of the deeply non-perturbative
domain of strong interactions.

\end{document}